\documentclass[12pt,preprint]{aastex}
\usepackage{graphicx,epsfig}
\begin{document}

\title{Galactic Positrons From Localized Sources}
\author{David Eichler\altaffilmark{1}, Irit Maor\altaffilmark{2}}
\altaffiltext{1}{Physics Department, Ben-Gurion University,
Beer-Sheva, Israel, {\it{eichler@bgumail.bgu.ac.il}}}
\altaffiltext{2}{Institute of Astronomy, University of Cambridge,
Madingley Road, Cambridge CB3 0WA, UK,
{\it{i.maor@damtp.cam.ac.uk}}}
\begin{abstract}
The anomalous bump in the cosmic ray positron to electron ratio at
$10~GeV$ can be explained as being a component from a point source
that was originally harder than the primary electron background
and degrades  due to synchrotron and inverse Compton losses in the
Galaxy while propagating to the Earth's vicinity. The fit is
better than can be obtained with  homogeneous injection and is
attributed to a minimum age threshold. Annihilating neutralinos
can provide a fair fit to the data if they have a mass just  above
1/2 the mass of the $Z^o$ and if they annihilate primarily in
distant density concentrations in the Galaxy. A possible
observational consequence of this scenario would be intense
inverse Comptonization of starlight at the Galactic center, with a
sharp energy cutoff in the emergent photons as a possible
signature of the neutralino mass.
\end{abstract}


\maketitle

\section{Introduction}
The possibility that weakly interacting dark matter particles
(WIMP's) could annihilate into detectable cosmic radiation was
suggested by  \citet{Si84}. \citet{Ty87} noted a reported positron
excess, curiously localized near $10~ GeV$
[\citet{Mu85,mt,data,B98}] and considered whether it could be due
to the annihilation of photinos (as a simple example of
neutralinos) in the tens of $GeV$ mass range. The difficulty was
that this process, given the laboratory constraints on the
neutralinos, seemed to fall short of providing enough positrons,
and the results were not published. Various papers on this excess
eventually appeared [\citet{t2,e2,turner,coutu}], and some noted
that the potential for positron excess could be bolstered by
clumpiness in the annihilating dark matter or by decay of weakly
unstable dark matter particles.

The approach usually found in present literature is to try and fit
the overall $e^+/(e^++e^-)$ ratio, without giving special
attention to the curious behaviour at $10~GeV$, see
\citet{Hooper:2004bq}. \citet{Ba01} considered a whole class of
minimal standard supersymmetric models and failed to get any
non-monotonicity in the $e^+/(e^++e^-)$ ratio. \citet{p1}
(henceforth Paper I) considered annihilation of particles through
the $Z^o$-channel (which we shall henceforth refer to as virtual
$Z^o$-decay) noted that non-relativistic virtual $Z^o$ decay (i.e.
when the rest mass of the annihilating dark matter particle is
slightly above 1/2 the $Z^o$ mass) provide a remarkably good fit
to the observed $e^+/(e^++e^-)$ ratio below $10~GeV$ mainly due to
the positrons that emerge from decaying muons. At higher energies,
however, the predicted $e^+/(e^++e^-)$ ratio rose above the
observed values within conventional assumptions about the
injection and propagation. In particular, it was assumed in Paper
I that the positrons and primary electrons are each injected with
the same spatial profile, and that their propagation in the Galaxy
is identical. The reason for this rise is that some $Z^o$'s decay
directly into high energy $e^+e^-$ pairs so that the $e^+$ energy
is half the $Z^o$ mass, and this gives rise to a high energy bump
in the $e^+/(e^++e^-)$ ratio at about $50~GeV$. While this bump
can be partially washed out by losses and escape, it was found
that the high energy $e^+/(e^++e^-)$ ratio is nevertheless
apparently too high to fit the observations to within $1\sigma$
error bars.  As discussed in Eichler (1989) this is a generic
problem for any positron source that is significantly harder than
the primary electrons above $10~GeV$.

However, the dark matter annihilation  scenario for explaining the
positron excess in any case requires that clumping of the dark
matter, and a likely place for this is near the Galactic center.
This means that the positrons in our neighborhood that are dark
matter annihilation products would have a minimum age, i.e. the
time needed to diffuse from the source to our  neighborhood, and
the age distribution of the positrons that make it to the Earth's
vicinity contain fewer young positrons than the age distribution
that one associates with the standard leaky box model. In this
letter we consider that the positrons are injected by an
effectively point source at a finite distance, and show that it
greatly improves the fit over that obtained in Paper I. The
positron bump at $\sim 10~GeV$ can be attributed to halo-type age
$\sim 3 \times 10^7~yr$ for the positrons, for over such a
lifetime, positrons losing energy by synchrotron and inverse
Comptonization would end up at about this energy.

We will find  that obtaining a good fit from a single source with
a single diffusion coefficient is difficult. However, it is well
known that below a certain intensity,  e.g. far enough up front in
a diffusion front,  cosmic rays can freely stream. Such behavior
is observed upstream of the Earth's bow shock. Theoretical reasons
for such free streaming include the difficulty of resonantly
scattering cosmic rays through 90 degree pitch angle in the linear
wave amplitude regime. Also, a larger, counterstreaming , finitely
stable component of cosmic rays would stabilize the smaller   free
streaming positron component. We therefore allow for the
possibility that a small fraction of the positrons freely stream,
and arrive at the Earth's vicinity much younger than the rest. We
find that this improves the fit still further. We find that the
low energy non-monotonicity, which appears in the {\it injection}
spectrum from non-relativistic $Z^o$ decay Paper I, can be
produced also by using a simple power law for the injected
spectrum (using the same propagation model).  It can thus be
produced by the combination of a harder (but monotonic) spectrum
of injected positrons and propagation effects.  We conclude that
it is still too early to unambiguously  interpret the low energy
behavior of the spectrum as a signature of self annihilating dark
matter.

%%%%%%%%%%%%%%%%%%%%%%%%%%%%%%%%%%%%%%%%%%%%%%%%%%%%%%%%%%%%%%%%%%%%
\section{Equations and Results}

The steady state diffused equation for the particle number
density, $n(x,r)$, is
\begin{eqnarray}
 \frac{\partial n}{\partial t} & = & 0=\hat{D}n
        -Rn+\frac{1}{m_Z}\frac{\partial}{\partial x}\left(m_Z\frac{dx}{dt}n \right)
        +I(x)\delta\left(r/L\right) \label{eq}
\end{eqnarray}
$x=E/m_Z$, $\hat{D}$ is a diffusion operator, $R=B{x}^{0.5}$ is
the escape rate with $B\sim few\times 10^{-15}1/s$,
$m_Z\frac{dx}{dt}=Ax^2$ with $A=8.5\times 10^{-16}~erg/s$ is the
Compton loss rate, corresponding to an electromagnetic energy
density in the Galaxy of $10^{-12}~erg/cm^3$, $L$ is a distance
scale, and $I(x)$ is the spectrum injected by a point source.

We assume as in Paper I that  the primary electrons and background
positrons are injected homogeneously, $\hat{D}n_b=0$:
\begin{eqnarray}
 n_b(x) & = & \frac{m_Z}{Ax^2}exp\left[-\frac{2m_ZB}{A\sqrt{x}} \right]
        \int_x^{\infty}I_b(x')exp\left[
        \frac{2m_ZB}{A\sqrt{x'}} \right]dx' \\
 I_{b,e^-}(x) & = & Cx^{-2} ~~~~~~for~~background~~e^- \nonumber \\
 I_{b,e^+}(x) & = & Dx^{-2.8} ~~~~for~~ background~~e^+
\end{eqnarray}

For the $Z^o$ decay injected spectrum we take the diffusion to be
one-dimensional with a diffusion coefficient ${\mathcal{D}}$,
$\hat{D}={\mathcal{D}}\frac{\partial^2}{\partial x^2}$, and with
boundary conditions such that $\frac{\partial n}{\partial
r}|_{r=L}=0$ (conserving the number of particles except for the
escape term). With these boundary conditions, the solution to eq.
(\ref{eq}) is
\begin{eqnarray}
 n_Z(x,r) & = & \frac{m_Z}{Ax^2}exp\left[-\frac{2m_ZB}{A\sqrt{x}} \right]
        \times \nonumber \\
 &   &  \int_x^{\infty}I_Z(x')exp\left[\frac{2m_ZB}{A\sqrt{x'}} \right]
        \sum_{-\infty}^{\infty}\cos\left[\frac{\pi n r}{L} \right]
        exp\left[-\left(\frac{\pi n}{L} \right)^2\frac{m_Z{\mathcal{D}}}{A}
        \left(\frac{1}{x}-\frac{1}{x'} \right)\right]dx'
\end{eqnarray}
While ${\mathcal{D}}$ (the diffusion coefficient) and $L$ (the
size of the leaky box) are free parameters, we took $r=8~Kpc$, the
distance to the galactic centre. $K\equiv{\cal{D}}/r^2$ gives the
inverse
time for diffusion. \\
We have chosen a one dimensional diffusion because it gives
somewhat better results than 3 dimensional diffusion. This is
physically plausible if one considers magnetic fields which will
confine the movement of the charged particles. So the geometry is
tube-like, with an effective cross section such that the total
volume is the galactic volume, $(20~kpc)^3$.

$I_Z(x)$ is the $Z$ decay products,
\begin{eqnarray}
  I_Z(x) & = & N\left(\frac{}{}0.0344I_{e}(x)+0.0344I_{\mu}(x)+
            0.0069I_{\tau}(x)+0.6916I_h(x)\frac{}{}\right) \\
  &  &  \nonumber \\
  I_e(x)& = & \delta\left(x-\frac{1}{2}\right) \nonumber \\
  I_{\mu} (x)& = & \frac{2}{3}
            \left(\frac{}{}5-36x^2+32x^3\frac{}{} \right) \nonumber \\
  I_{\tau}(x)& = &\frac{2}{3}
            \left(\frac{}{}5-36x^2+32x^3\frac{}{} \right) \nonumber \\
  & + &\frac{2}{9}\left[-\frac{95}{3}-108x^2+\frac{1408}{3}x^3
        -\left(25+324x^2+128x^3\right)\ln \left(2x \right) \right] \nonumber \\
  I_{h}(x) & = & \frac{14}{9}\int_{\frac{28}{9}x}^1
        \frac{d\bar{x}}{\bar{x}}
        10^{a_k-b_k\bar{x}} \nonumber \\
    &  & a_1=3,~b_1=10~~~~~~~0<\bar{x}<0.1  \nonumber \\
    &  & a_2=2,~b_2=4 ~~~~~~0.1<\bar{x}<1  \nonumber
\label{is}
\end{eqnarray}
Each $I_{ch}$ describes the $ch$ channel of decay, and the
pre-factors correspond to the branching ratios. The calculation
was done in zeroth order, assuming 3 massless families and
neglecting the top quark, for details see Paper I. Following the
discussion there, we take $N=1.3\times 10^{-29}~1/(cm^3~s)$ as the
annihilation rate per unit volume.
\\

%%%%%%%%%%%%%%%%%%%%%%%%%%%%%%%%%%%%%%%%%%%%%%%%%%%%%%%%%%%%%%%%%%%%

\begin{figure}
% \title{}
 \begin{center}
 \epsfig{file=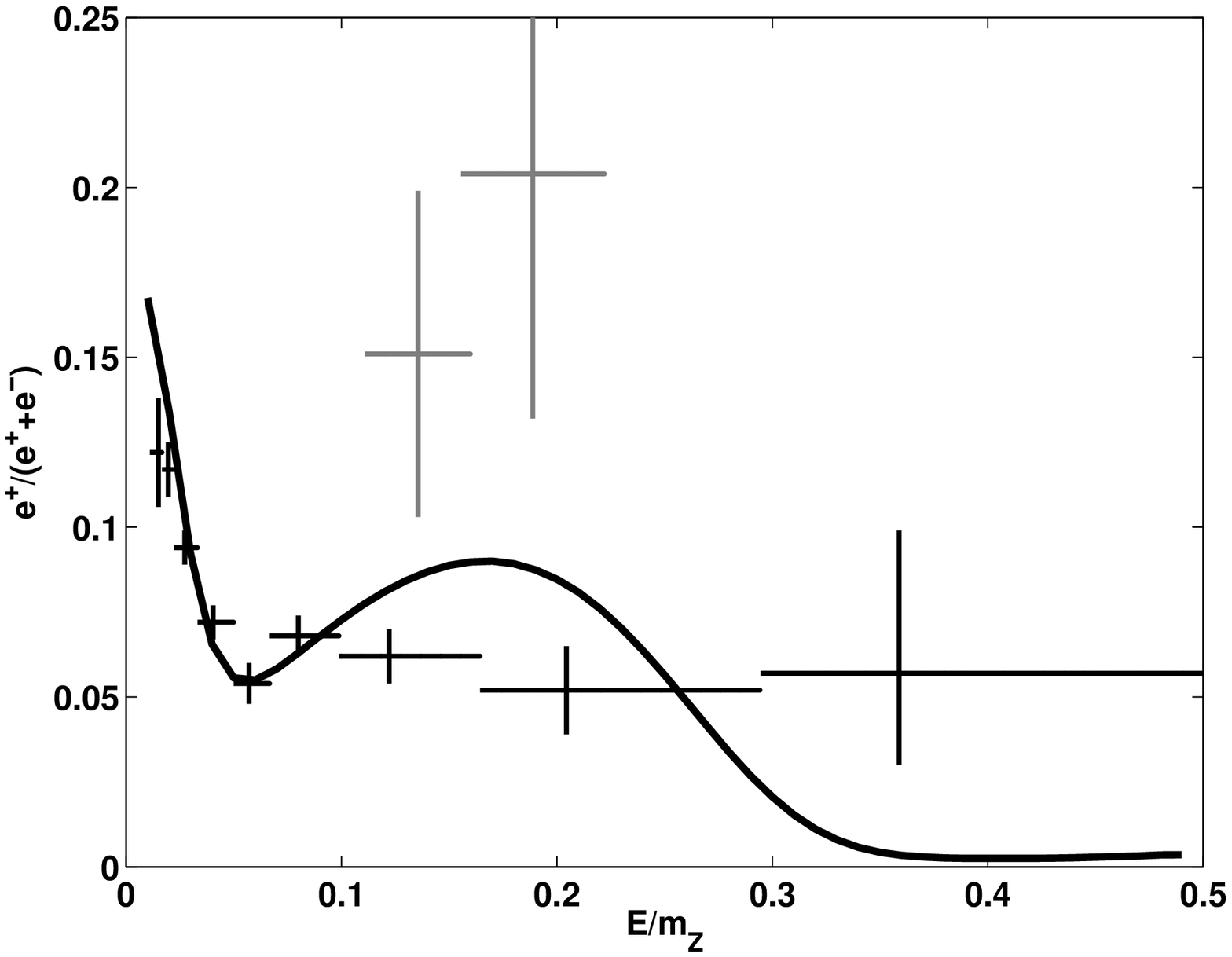,height=70mm}
 \end{center}
 \caption{The $e^+/(e^++e^-)$ as a function of $x=E/m_Z$, for a
            single source $Z$ decay injected spectrum.
            $A=8.5\times 10^{-16}~\frac{erg}{s}$,
            $B=7.1\times 10^{-15}~\frac{1}{s}$,
            $C=4.0\times 10^{-29}~\frac{1}{cm^3~s}$,
            $D=1.3\times 10^{-31}~\frac{1}{cm^3~s}$,
            and $K=1.9\times 10^{-16}~\frac{1}{s}$.
            Data taken from \citet{data} (black) and \citet{mt}
            (grey).
 }
 \label{singleZ}
\end{figure}

\begin{figure}
% \title{}
 \begin{center}
 \epsfig{file=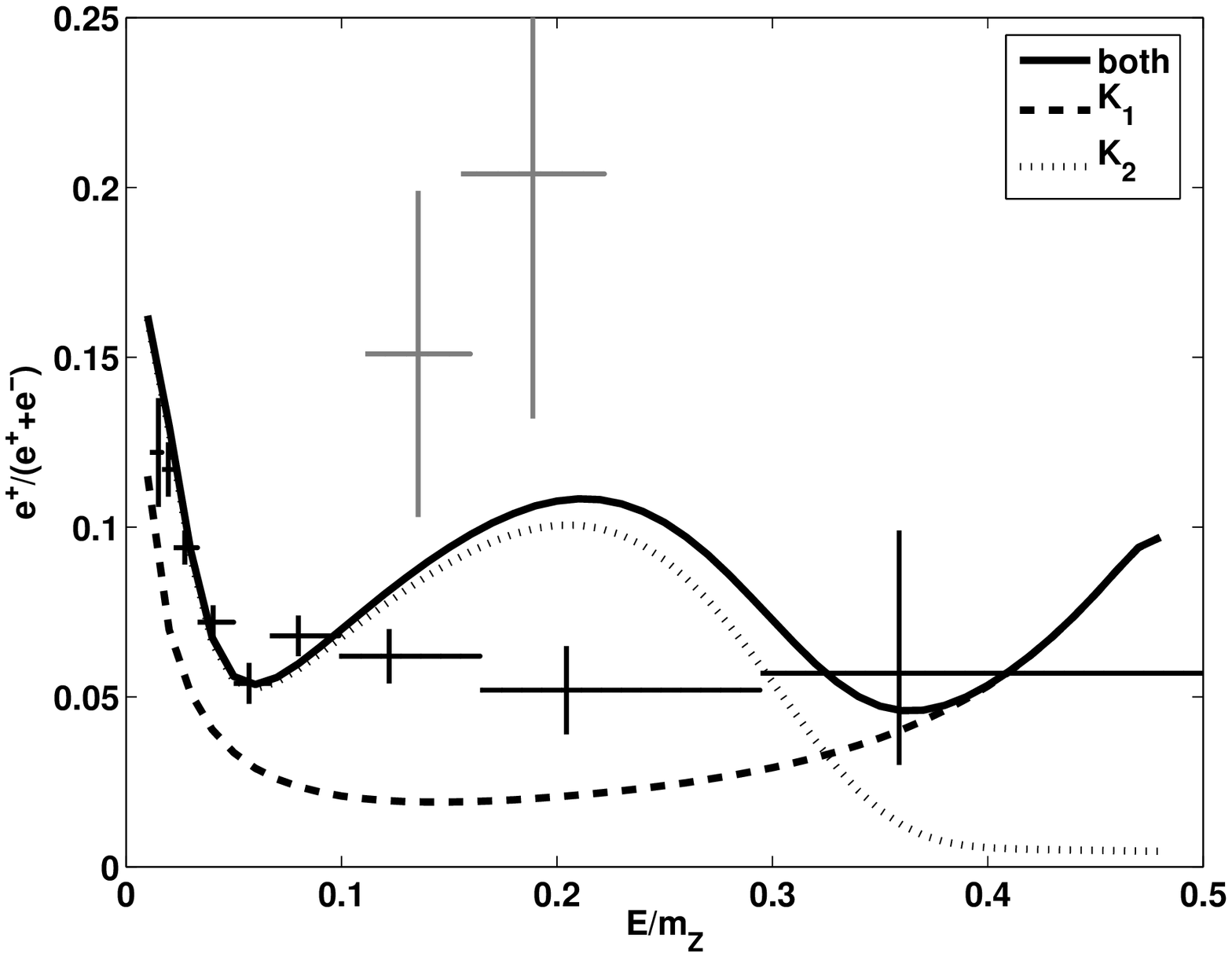,height=70mm}
 \end{center}
 \caption{The $e^+/(e^++e^-)$ as a function of $x=E/m_Z$, for a
            combination of 2 sources of $Z$ decay injected spectrum.
            $A=8.5\times 10^{-16}~\frac{erg}{s}$,
            $B=7.6\times 10^{-15}~\frac{1}{s}$,
            $C=4.9\times 10^{-29}~\frac{1}{cm^3~s}$,
            $D=1.3\times 10^{-31}~\frac{1}{cm^3~s}$,
            $K_1=2.8\times 10^{-14}~\frac{1}{s}$
            and $K_2=2.8\times 10^{-16}~\frac{1}{s}$. The
            ratio between the two components is $1:5$.
            Data taken from \citet{data} (black) and \citet{mt}
            (grey).
 }
 \label{combo}
\end{figure}

Fig. (\ref{singleZ}) shows a fit with a single point source of
$Z^o$ decay and a single diffusion coefficient. The good fit to
the low energies from Paper I is still present, but at the price
that the excess in energies toward $x=1/2$ is now is suppressed by
the finite age effect. As the figure shows, we are now facing a
scenario which is opposite to Paper I; the finite age effect
tends to suppress the high energy excess at the price of killing
it off altogether.\\
However, there are several possibilities that avoid this problem:
There may be more than one source, and there may be more than one
route (roughly guided by magnetic field lines) by which the
particles diffuse or freely stream from the source to our
vicinity.  High energy particles diffuse much less than low energy
ones because they are fewer in number and create less waves. So
their self-generated scattering is less efficient. Thus, the
fraction of free streaming particles should be higher at higher
energy. Fig. (\ref{combo}) shows a combination of two $Z^o$ decay
components, an older, larger one that arrives via diffusion, and a
younger, smaller component that has managed more free streaming.
This figure illustrates that if one takes an age distribution into
account, the flexibility in adjusting the high energy spectrum
becomes much larger, and can be fitted to the data.

\begin{figure}
 \title{}
 \begin{center}
 \epsfig{file=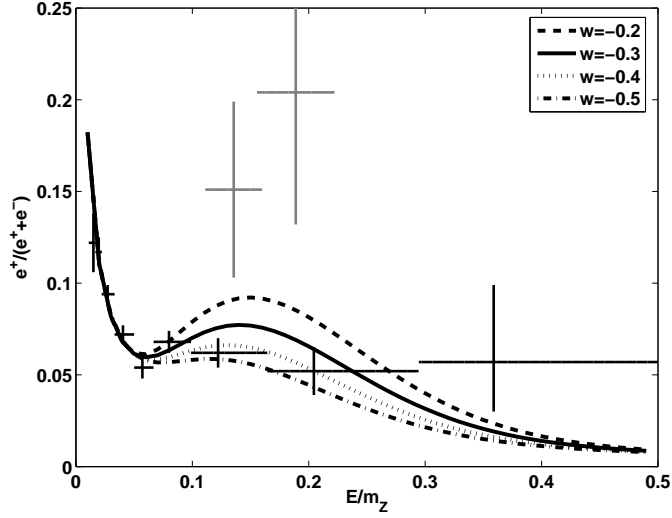,height=70mm}
 \end{center}
 \caption{The $e^+/(e^++e^-)$ as a function of $x=E/m_Z$, for
            various power laws, $Nx^{w}$, as the injected spectrum.
            $A=8.5\times 10^{-16}~\frac{erg}{s}$,
            $B=4.4\times 10^{-15}~\frac{1}{s}$,
            $C=1.7\times 10^{-29}~\frac{1}{cm^3~s}$,
            $D=1.1\times 10^{-31}~\frac{1}{cm^3~s}$,
            and $K=6.6\times 10^{-17}~\frac{1}{s}$.
            Data taken from \citet{data} (black) and \citet{mt}
            (grey).}
 \label{mpl}
\end{figure}

\begin{figure}
% \title{}
 \begin{center}
 \epsfig{file=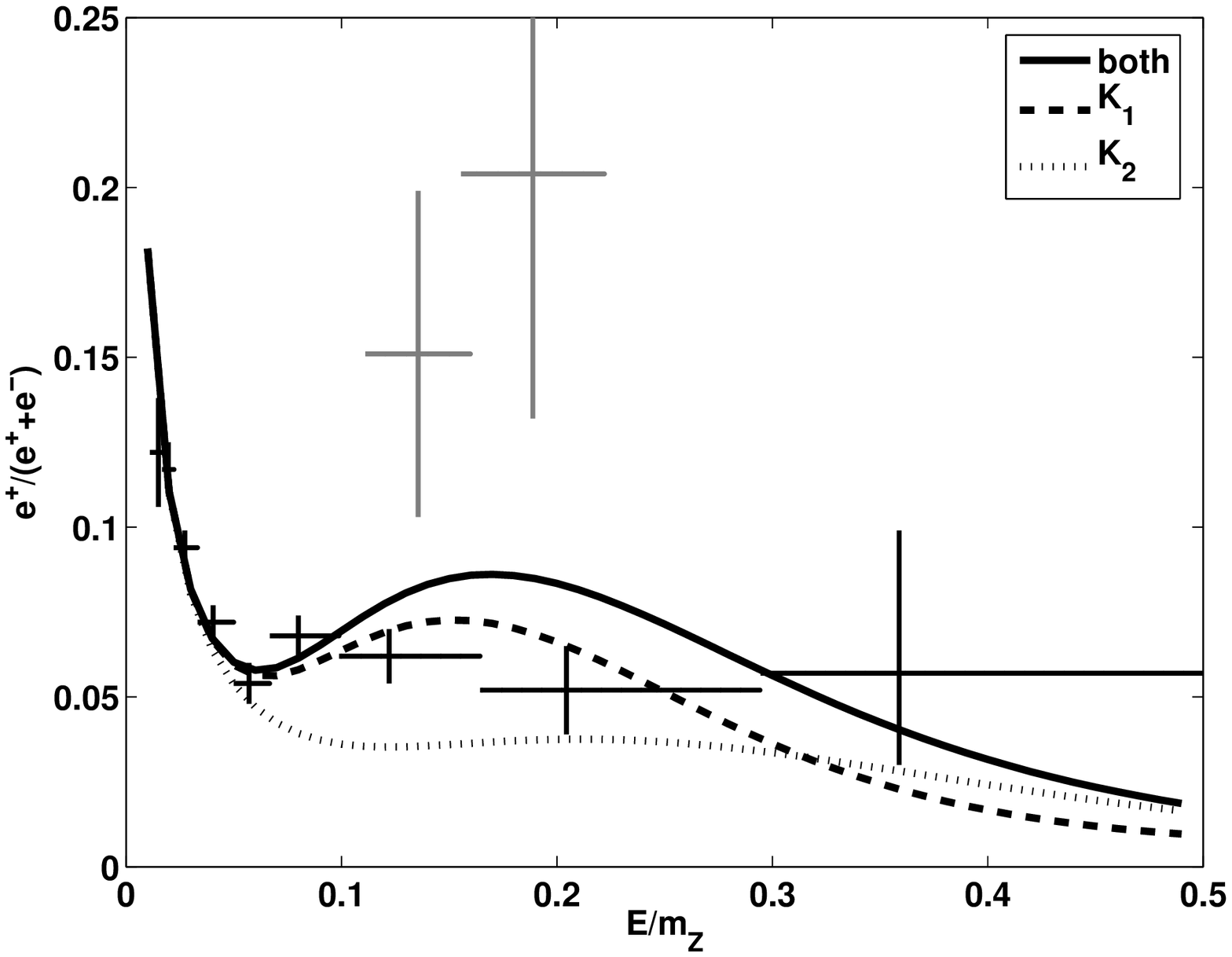,height=70mm}
 \end{center}
 \caption{The $e^+/(e^++e^-)$ as a function of $x=E/m_Z$, for a
            combination of 2 sources of power law ($w=-0.3$)
            injected spectrum.
            $A=8.5\times 10^{-16}~\frac{erg}{s}$,
            $B=4.4\times 10^{-15}~\frac{1}{s}$,
            $C=1.7\times 10^{-29}~\frac{1}{cm^3~s}$,
            $D=1.1\times 10^{-31}~\frac{1}{cm^3~s}$,
            $K_1=7.2\times 10^{-17}~\frac{1}{s}$
            and $K_2=1.2\times 10^{-16}~\frac{1}{s}$. The
            ratio between the two components is $10:1$.
            Data taken from \citet{data} (black) and \citet{mt}
            (grey).
 }
 \label{plcombo}
\end{figure}

For sake of comparison, we also include a power law injected
spectrum, fig. (\ref{mpl}) shows various power laws, and fig.
(\ref{plcombo}) shows two components with different ages. We find
that as long as the injected power law is hard enough, one can
produce a low energy ($5-10~ GeV$) dip. The quality of the fit is
almost as good for a power law as for virtual $Z^o$ decay. We
consider the low energy dip to have qualitatively more
significance than the higher points and have emphasized those data
points accordingly in choosing the best fit. We have deliberately
not quantified this with the standard statistical measures. Trying
to get the statistically best parameters (for either power law or
$Z^o$ decay as injected spectrums) would wash out the low energy
behavior that we are
focusing on.  \\

Although we can reproduce the $7~GeV$ dip, the peak at $E\sim 15~
GeV$ is still too big for the HEAT data (though too small for the
earlier data). This seems to be a generic feature of our results,
regardless of whether the injection source is virtual $Z^o$ decay
or a power law. The problem would be worse if the virtual $Z^o$
had an energy well above $m_Z$.

\section{Possible Observational Consequences}

The hypothesis that the neutralino mass $m_{\chi}$ is only
slightly more than half the $Z^o$ mass is motivated by several
factors: The annihilation cross section can be resonantly enhanced
by a factor of 2 or 3 more than that during annihilation in the
early universe, when the thermal broadening of the Z resonance
somewhat exceeded its natural width, \citet{gs91}. Moreover,
assuming the smallest allowable mass allows the greatest
annihilation rate since the annihilation reaction rate is fixed by
the condition that it allows a given cosmic dark matter
contribution. (Although dark matter clumping can enhance the
annihilation rate, a plausible level of such enhancement is
limited by observational constraints on dark matter clumping that
are set by stellar distributions in galactic centers.) Making
$m_{\chi}$ just above $m_Z$ causes the $Z^o$ resonance to be
asymmetric, but this would be hard to measure experimentally
because of the weak coupling of the emerging neutralinos at CM
collision energies above $2m_{\chi}$. On the other hand, that  the
annihilation cross the virtual $Z^o$ must be close to its mass
shell if it is to provide a decent fit suggests that its loop
corrections would be large and it might be discernable or
falsifiable with particle collider data on processes that depend
on such loop corrections.\\

In an astronomical context, a possible observational consequence
of a point source of positrons at the Galactic center could be
inverse Comptonization of  starlight, which is far more intense
than at a typical point in the Galaxy. The profile of Galactic
starlight near the Galactic center is given by \citet{kent}. The
derived photon energy density is then
$U(r)=4.3\times10^{-9}(r/pc)^{-0.85}~erg/s$. Assuming the
positrons are produced within a typical radius r of the Galactic
Center, they produce a minimum of $E_{IC}(\gamma) \equiv
\int_r^{100pc}\gamma^2 \sigma_T U(r)dr$ in inverse Compton (IC)
scattered starlight before escaping the central $100~pc$ region,
and the luminosity  $L(\ge \gamma m_ec^2)$ above $\gamma^2
\epsilon_{ph}$, where $\epsilon_{ph}$ is the typical energy of the
pre-scattered starlight photons, is
$\int_{\gamma}^{\infty}I(\gamma' m_ec^2)E_{IC}(\gamma')d\gamma'$.
The most energetic $e^+e^-$ pairs alone, that is those which
result directly from the $Z^o$ decay ($I_e(x)$ in eq. \ref{is}),
will produce IC luminosity of $3.8\times 10^{35}~erg/s$.

\begin{figure}
% \title{}
 \begin{center}
 \epsfig{file=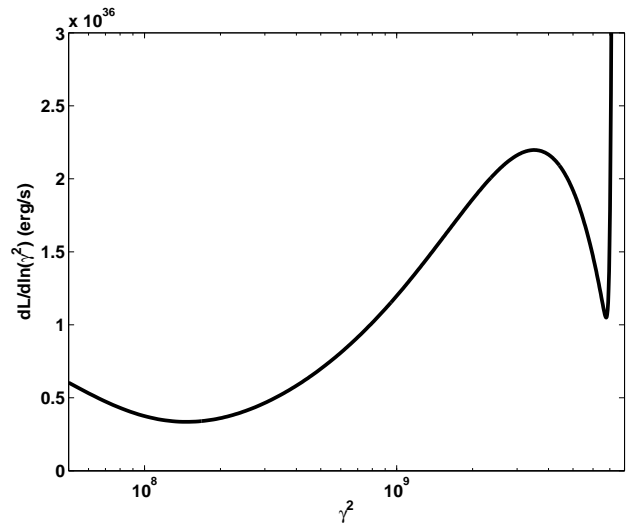,height=70mm}
 \end{center}
 \caption{ Differential IC luminosity due to the positrons only,
            $-dL/d\ln(\gamma^2)\times (20~kpc)^3/V$
            as a function of $\gamma^2$.
 }
 \label{lnIC}
\end{figure}

Fig. (\ref{lnIC}) shows the logarithmic derivative of the IC
luminosity due to the positrons {\it{only}}, $-\gamma^2
L_{\gamma^2} = -dL/d\ln(\gamma^2)$, as a function of the square of
the positron Lorentz factor, $\gamma^2$. Shown in the figure is
the minimum IC luminosity as a function of the frequency scaled to
the frequency of the pre-scattered photon (x axis).  The minimum
luminosity assumes that the positrons emerge from the central
region in a straight line. If the mean free path $\lambda$ is less
than $100~ pc$, then the predicted luminosity goes up by roughly a
factor of (100pc/$\lambda$). \\

For dark matter annihilation that yields direct monochromatic $e^+
e^-$ pairs at Lorentz factor $\gamma_o$, this would translate into
a sharp cutoff in the IC gamma rays of $\gamma_o^2\epsilon_{ph}$
or about $3\gamma_o^2~eV$. In our particular example
$m_{\chi}\simeq m_Z/2$, this would lead to a cutoff at
$10^{10}\epsilon_{ph}\sim 30~GeV$. This would in principle be
detectable by MAGIC, see for example \citet{magic}, if the
location were suitable for observing the Galactic center.
Alternatively, it could be detected by HESS if the energy
threshold could be pushed to below $30 ~GeV$. This scenario would
{\it not} explain the $TeV$ photons from the Galactic center
recently reported by the HESS collaboration, \citet{hess}. If,
however, there is annihilation in the Galactic Center of heavier
dark matter particles, then direct $e^+ e^-$ pairs might be
detectable via such a cutoff in the $TeV$ gamma ray
spectrum at $m_{\chi}/2$.\\

In conclusion, we find that the cosmic ray positron data can be
fit with more than one hard source of positrons provided that a)
they have a chance to lose energy before escaping the Galaxy and
b) they have a minimum age (e.g. they come from discrete, distant
sources), unlike the background primary electrons.  They need not
be from dark matter annihilation, but a best case scenario for
this is not confidently ruled out by existing data. Detection of
inverse Compton radiation with good energy resolution can in
principle provide information as to the spectrum of the positrons
and their point of origin.

\acknowledgments{We thank T. Alexander, E. Baltz, and A. Dekel for
helpful discussions. DE gratefully acknowledges the support from
the Israel-U.S. Binational Science Foundation, the Israel Science
Foundation and the Arnow Chair of Physics at Ben Gurion
University. IM gratefully acknowledges the support from the
Leverhulme Trust.}

%%%%%%%%%%%%%%%%%%%%%%%%%%%%%%%%%%%%%%%%%%%%%%%%%%%%%%%%%%%%%%%%%%%%

%%%%%%%%%%%%%%%%%%%%%%%%%%%%%%%%%%%%%%%%%%%%%%%%%%%%%%%%%%%%%%%%%5555
\newpage

%%%%%%%%%%%%%%%%%%%%%%%%%%%%%%%%%%%%%%%%%%%%%%%%%%%%%%%%%%%%%%%%

%\newpage

\end{document}